\documentclass[prb,twocolumn,showpacs,amsmath,amssymb,eqsecnum,
nobibnotes]{revtex4}
\usepackage{graphicx}
\usepackage{dcolumn}
\usepackage{bm}
\begin{document}
\title{Spin relaxation in a Rashba semiconductor in an electric field}
\author{O.Bleibaum}
\email{olaf.bleibaum@physik.uni-magdeburg.de}
\affiliation{Institut f\"ur Theoretische Physik, Otto-von-Guericke
Universit\"at Magdeburg, PF4120, 39016 Magdeburg, Germany}
\begin{abstract}
The impact of an external electric field on the spin relaxation in a 
disordered, two-dimensional electron system is studied within the framework 
of a field-theoretical formulation. Generalized Bloch-equations for 
the diffusion and the decay of an initial magnetization are obtained.
The equations are applied to the investigation of spin relaxation 
processes in an electric field. 
\end{abstract}
\maketitle
\section{Introduction}
In recent years much attention has been paid to the emerging field
of spintronics in semiconductor physics. Much of the interest in this 
field is stimulated by the spin-field-effect transistor proposed by 
Datta and Das \cite{Datta}. In this proposal the Rashba spin-orbit 
interaction is used to manipulate the charge transfer through
a transistor like device by controlling the electron spin. The Rashba
interaction, which is the conventional spin-orbit interaction with
a constant  electric field, affects the electron spin like a momentum
dependent magnetic field.

In practice the realization of the proposal by Datta and Das has proven 
to be a challenging task. In particular the injection of spins into 
non-magnetic semiconductors has turned out to be demanding, although
successful injection has been reported by several authors \cite{Schmidt1,
Awschalom,Ploog,Schmidt2,Johnson}. Despite  this fact the importance 
of the Rashba interaction for the concept of the spin-field effect 
transistor has motivated further studies on its impact on charge and spin 
transport properties. Particular attention has been paid, e.g., 
to investigations of spin-accumulations\cite{Molenkamp1} and to the
spin-galvanic effect\cite{Ganichev}, to an intrinsic spin-Hall 
effect\cite{McDonald}, and to the weak localization-antilocalization 
transition in the conductivity \cite{Aleiner,Brouwer1,Zumbuhl,Koga,Miller}.

Since in the Rashba model the electron wave-vector is coupled to the spin
the  spin is quickly randomized by scattering events. Therefore,
the ballistic transport regime is the most important regime for the
spin-field effect transistor proposed by Datta and Das. Accordingly, most 
of the investigations focus on the situation in which the disorder energy
is small compared to the Rashba level splitting. Recently, however, a 
generalized 
spin-field effect transistor has been proposed\cite{Loss} which is also 
expected to work in the diffusive regime, even if the Rashba level 
splitting is small compared to the characteristic disorder energy. This raises
the question which impact of the Rashba interaction on the charge and
spin transport properties can be expected in this limit.

The aim of the paper is to provide a theory for the spin-relaxation 
in a a two-dimensional 
electron gas in the diffusive regime, in the limit that the characteristic
disorder energy is large  compared to the Rashba level splitting on a time 
scale,
in which inelastic processes can be ignored. To this end we 
generalize the non-nonlinear $\sigma$-model to include also the impact of the 
Rashba interaction and an in-plane electric field. Using the generalized 
$\sigma$-model we derive generalized Bloch-equation and use these 
equations for the investigation of an initial magnetization. 
\section{Basic equations}
We consider charge carriers in a two-dimensional plane. The motion of 
the charge carriers is described by the Hamilton operator
\begin{equation}\label{B1}
H=\frac{{\hat{\bm p}}^2}{2m}+{\bm F}\cdot{\bm x}
+{\bm\sigma}\cdot({\bm N}\times{\hat{\bm p}})+V({\bm x}).
\end{equation}
Here ${\hat{\bm p}}$ is the momentum operator, $m$ is the effective mass,
${\bm F}$ is a constant field in the two-dimensional plane,   
${\bm N}$ is a constant vector perpendicular to the plane, and $V({\bm x})$
is a random potential. The components
of the vector ${\bm \sigma}=\sigma_x{\bm e}_x+\sigma_y{\bm e}_y+
\sigma_z{\bm e}_z$ are the Pauli matrices. ${\bm e}_x$ and ${\bm e}_y$
are the unit vectors in the two-dimensional plane and ${\bm e}_z$
is a unit vector transverse to the plane. Accordingly, 
${\bm N}=N_z{\bm e}_z$ and ${\bm F}=F{\bm e}_x$. The random potential
is characterized by a Gaussian distribution function with zero mean and
second moment
\begin{equation}\label{B2}
\langle V({\bm x})V({\bm x'})\rangle_{dis}=\frac{1}{\pi\nu\tau}\delta
({\bm x}-{\bm x'}).
\end{equation}
Here $\langle\dots\rangle_{dis}$ denotes the configuration average, $\nu$ is 
the density of states per spin at the Fermi-level for $N_z=0$, and $\tau$ is 
the single-particle scattering time.

In our investigation we focus on configuration averages of retarded and 
advanced Green functions. These functions are solutions to the differential
equation
\begin{equation}\label{B3}
((\pm i\omega+E){\bm 1}-H)G^{R/A}({\bm x},{\bm x'}|E,\omega)=
{\bm 1}\delta({\bm x}-{\bm x'}).
\end{equation}
In this equation $G^{R/A}$ is a $2\times 2$ matrix in spin-space ,
${\bm 1}$ is the $2\times 2$ unit matrix and $E$ and $\omega$ are
fixed energies. As a boundary condition we require 
that the Green functions vanish at infinity. In this case the configuration
averaged Green functions are invariant against translations by a fixed vector 
${\bf a}$ if the energy $E$ is also shifted by ${\bm F}\cdot {\bm a}$.
Accordingly, they satisfy the relationship
\begin{eqnarray}\label{B4}
\langle G^{R/A}\rangle_{dis}&(&{\bm x}+{\bm a}, {\bm x'}+{\bm a}|E,\omega) 
\nonumber\\
& &=\langle G^{R/A}\rangle_{dis}
({\bm x}, {\bm x'}|E-{\bm F}\cdot{\bm a},\omega).
\end{eqnarray}
To calculate the configuration averages of the Green functions we use the 
field theoretical formulation of the Refs.
[\onlinecite{McKane,Pruisken,Belitz1}].  
In this formulation the replica trick is used to derive an effective action 
for the calculation of the configuration averaged Green functions. Following
the standard derivation we find that for the problem at hand the partition
function takes the form
\begin{equation}\label{B5}
Z=\int DQ\exp(A_Q),
\end{equation}
where
\begin{equation}\label{B6}
A_Q=-\frac{\pi\nu\hbar}{8\tau}{\mbox{ tr}}Q^2
+\frac{1}{2}{\mbox{tr}}\ln G_Q^{-1}
\end{equation}
Here $Q$ is a hermitian $(2n_r+4)\times(2n_r+4)$ matrix field ($n_r$-number of
replica copies+ 2 components particle-hole space+ 2 components spin indices) 
with components $Q_{\alpha m\;\alpha'n}^{ij}({\bm x})$ ($\alpha,\alpha'$-
replica indices, $i,j$- particle-hole indices, $m,n$-spin indices).
It satisfies the relationship $QC=CQ^T$, where $Q^T$ is the transposed
matrix and
\begin{equation}\label{B6a}
C=\left(
\begin{array}{cc}
0&i\sigma_y\\
i\sigma_y&0
\end{array}
\right).
\end{equation}
The tr in Eq.(\ref{B6}) contains both the summation over the discrete 
indices and the integration over space.
Furthermore,
%
%
\begin{eqnarray}\label{B7}
{G_Q^{-1}}_{\alpha m\;\alpha'n}^{ij}({\bm x},{\bm x'})=
&[&\hspace{-0.5em}\{(i\omega_{\alpha}+E)\delta_{mn}
-H_{mn}\}\delta_{\alpha\alpha'}\delta_{ij}\nonumber\\
&+&i\frac{\hbar}{2\tau}
Q_{\alpha m\;\alpha'n}^{ij}({\bm x})]\delta({\bm x}-{\bm x'}),
\end{eqnarray}
%
%
where $\omega_{\alpha}=\omega\Lambda_{\alpha}$, $\Lambda_{\alpha}=1$ 
for $\alpha>0$ and $\Lambda_{\alpha}=-1$
for $\alpha\leq 0$. The index $\alpha$ can take on values between
$-(n_r-1)$ and $n_r$. It is understood that the limit $n_r\to 0$ is 
taken at the end of the calculation.
\section{The saddle-point Green functions}  
To calculate the configuration-averaged Green function in saddle-point 
approximation we now look on the saddle-point solution of the effective
action $A_Q$, defined by 
\begin{equation}\label{SP1}
\frac{\delta A_Q}{\delta Q}|_{Q=Q_{SP}}=0.
\end{equation}
The saddle-point values of $Q$ and  $G_Q$ are diagonal with respect to the
replica indices, as are the exact expection values.  
Accordingly, ${Q_{SP}}_{\alpha m\,\alpha' n}^{\;ij}=
{Q_{\alpha}}_{mn}^{\;ij}\delta_{\alpha\alpha'}$ and 
${G_Q^{SP}}_{\alpha m\;\alpha' n}^{\;\;ij}={G_{\alpha}}_{mn}^{ij}
\delta_{\alpha\alpha'}$. From the Eqs. (\ref{B4}) and (\ref{B6}) we find
\begin{eqnarray}\label{SP2}
{Q_{\alpha}}_{mn}^{\;ij}({\bf x})&=&\frac{i}{\pi\nu}
{G_{\alpha}}_{mn}^{ij}(0,0|E-{\bm F}\cdot{\bm x},\omega)\nonumber\\
&\equiv&\tilde{Q}_{\alpha\;mn}^{\;\;\;\;ij}(\mu_x).
\end{eqnarray}
Here $\mu_x=E-{\bm F}\cdot{\bm x}$.

To find an equation for the Green function we use the  gradient expansion
\cite{Kadanoff}. To this end we first introduce Wigner coordinates,
according to ${\bm R}=({\bm x}+{\bm x'})/2$ and ${\bm r}={\bm x}-{\bm x'}$,
and perform a Fourier transformation with respect to the relative 
coordinates. The Green function in the new
coordinates ${\tilde G}$ is related to the old Green function by the 
relationship
\begin{eqnarray}\label{SP3}
G_{\alpha\,mn}^{\;\;\;ij}\hspace{-0.5em}&(&\hspace{-0.5em}{\bm R}+{\bm r}/2,
{\bm R}-{\bm r}/2|E,\omega)\nonumber\\
&=&\int\frac{d{\bm k}}{(2\pi)^2}e^{i{\bm k}\cdot
{\bm r}}{\tilde{G}}_{\alpha\,mn}^{\;\;\;ij}({\bm R},{\bm k}|E,\omega).
\end{eqnarray}
In calculating ${\tilde G}$ we restrict the consideration to the
lowest order in the gradient expansion. In this approximation the equation 
for the Green function takes the simple form
\begin{widetext}
\begin{eqnarray}\label{SP4}
\sum_j([(i\omega_\alpha+\mu_R-\epsilon_k){\bm 1}-
\hbar{\bm\sigma}\cdot({\bm N}\times{\bm k})]\delta_{ij}+i\frac{\hbar}{2\tau}
{\tilde Q}_{\alpha}^{ij}(\mu_R))
{\tilde G}_{\alpha}^{jk}({\bm R},
{\bm k}|E,\omega)
={\bm 1}\delta_{ik}
\end{eqnarray}
\end{widetext}
Here all objects are considered as $2\times2$ matrices in spin 
space, $\epsilon_k=\hbar^2k^2/2m$ and $\mu_R=\mu_x|_{{\bm x}={\bm R}}$. 

In order to solve the Eqs.(\ref{SP2})-(\ref{SP4}) self-consistently
we restrict the consideration to the classical accessible region
$\mu_R>0$. Furthermore, we consider only  the dirty limit, in which both 
the disorder energy $\hbar/2\tau$ and $\mu_R$ are large compared to the 
Rashba level splitting $|N_z|\sqrt{2m\mu_R}$ and assume that $\mu_R$  
is large compared to $\hbar/2\tau$. In this case we find that the 
self-consistent solution to the equations (\ref{SP2}) and (\ref{SP4})
takes the form
\begin{equation}\label{SP5}
{\tilde Q}_{\alpha\;mn}^{\;\;\;ij}(\mu_R)=\delta_{mn}\delta_{ij}
\mbox{sgn}\omega_{\alpha},
\end{equation}
\begin{equation}\label{SP6}
{\tilde{G}}_{\alpha\;mn}^{\;\;\;\;ij}({\bm R},{\bm k}|E,\omega)=
g_{\alpha\;mn}({\bm k}|\mu_R,\omega)\delta_{ij},
\end{equation}
\begin{widetext}
\begin{equation}\label{SP7}
g_{\alpha}({\bm k}|\mu_R,\omega)=
\frac{1}{M_{\alpha}(k|\mu_R,\omega)}
\left(
\begin{array}{cc}
i\omega_{\alpha}+\mu_R-\epsilon_k+i\frac{\hbar}{2\tau}\mbox{sgn}
\omega_{\alpha}& 
-iN_z\hbar k_{-}\\
iN_z\hbar k_+&i\omega_{\alpha}+\mu_R-\epsilon_k+i\frac{\hbar}{2\tau}
\mbox{sgn}\omega_{\alpha}
\end{array}\right),
\end{equation}

\begin{equation}\label{SP8}
M_{\alpha}({\bm k}|\mu_R,\omega)=(i(\omega_{\alpha}+\frac{\hbar}{2\tau}
\mbox{sgn}\omega_{\alpha})+\mu_R-E_+(k))
(i(\omega_{\alpha}+\frac{\hbar}{2\tau}
\mbox{sgn}\omega_{\alpha})+\mu_R-E_-(k)),
\end{equation}
\end{widetext}
where $k_{\pm}=k_x\pm i k_y$ and $E_{\pm}=\epsilon_k\pm \hbar |N_z| k$.
In writing down the saddle-point field (\ref{SP5}) we have ignored its 
imaginary part which only leads to an  $\alpha$-independent shift of 
$\mu_R$. We would like to note that the
simple structure of the saddle-point field is a consequence of the fact
that the density of states for $N_z=0$ is constant. In an energy 
dependent density of states the saddle-point would pick up an energy 
and field dependent contribution from the density of states, which 
would modify Eq.(\ref{SP5}).
\section{The non-linear $\sigma$-model}
The fluctuations around the saddle-point solution are characterized by
massive and soft modes. The latter describe the diffusion and spin relaxation
properties we are interested in. In the field-theoretical formulation 
the existence of these modes is a consequence of a continuous symmetry
of the effective action (\ref{B6}) which exists in the limit
$\omega, N_z\to 0$. In this limit the effective action is invariant against 
similarity transformations with hermitian matrices $B$, which satisfy the
relationship
\begin{equation}\label{SM1}
B^TCB=C.
\end{equation}
Such matrices  are generated by hermitian generators $W$ ($B=\exp(iW)$), 
which have the structure
\begin{eqnarray}\label{SM2}
W_{\alpha\alpha'}&=&\sum_{\lambda=0}^2[i{A_{\alpha\alpha'}}_0^{\lambda}
\sigma_{\lambda}\otimes\sigma_0+\sum_{i=1}^3
{S_{\alpha\alpha'}}_i^{\lambda}\sigma_{\lambda}\otimes\sigma_i]\nonumber\\
& &+{S_{\alpha\alpha'}}_0^3\sigma_3\otimes\sigma_0+i
{A_{\alpha\alpha'}}_i^3\sigma_3\otimes\sigma_i
\end{eqnarray}
Here ${S_{\alpha\alpha'}}_{\kappa}^{\lambda}$ 
(${A_{\alpha\alpha'}}_{\kappa}^{\lambda}$) are real matrices
symmetric (antisymmetric) in $\alpha$ and $\alpha'$, 
$\sigma_0={\bm 1}$, and $i=1,2,3$ corresponds to $x$, $y$ and $z$,
respectively. The components of the matrices $W_{\alpha\alpha'}$, which
are associated with the upper index of the matrices 
${S_{\alpha\alpha'}}_{\kappa}^{\lambda}$ and
${A_{\alpha\alpha'}}_{\kappa}^{\lambda}$ (that is with $\lambda$), 
correspond to the particle-hole indices. The lower indices characterize
the spin.  

In order to derive an effective action for the soft modes we follow the
standard derivation for the non-linear sigma-model 
\cite{McKane,Pruisken,Belitz1}. The complications originating from the
presence of the electric field we treat as in Ref.[\onlinecite{Bleibaum1}].
Doing so, we find that the effective action for the diffusion and spin 
relaxation modes takes the form
\begin{widetext}
\begin{equation}\label{SM3}
A_{\sigma}=\frac{\pi\nu\omega}{2}\int d{\bm x}\theta(\mu_x)\mbox{tr}(\Lambda
{\hat Q}({\bm x}))-\frac{\pi\nu\hbar}{8}\int d{\bm x}\theta(\mu_x)
D(\mu_x)\mbox{tr} 
({\bm D}{\hat Q}({\bm x})\cdot{\bm D}{\hat Q}({\bm x})).
\end{equation}
\end{widetext}
Here ${\hat Q}({\bm x})$ is the soft part of the matrix $Q({\bm x})$ 
generated from the saddle-point field $Q_{SP}({\bm x})$ by the local 
similarity transformation 
${\hat Q}({\bm x})=B({\bm x})Q_{SP}({\bm x})B^{-1}({\bm x})$,
\begin{equation}\label{SM4}
D(\mu_x)=\frac{\tau\mu_x}{m},
\end{equation}
and 
\begin{equation}\label{SM5}
{\bm D}{\hat Q}({\bm x})={\bm\nabla}{\hat Q}({\bm x})-i\frac{m}{\hbar}
[{\bm\sigma}\times{\bm N},{\hat Q}({\bm x})].
\end{equation}
The bracket $[\dots,\dots]$ symbolizes the commutator and the step function
$\theta(\mu_x)$ restricts the range of integration to the classical 
accessible region. The matrix ${\bm {\sigma\times N}}$ in Eq.(\ref{SM5})
is a matrix which only acts on the spin-indices but not on the replica
and the particle-hole indices.

In order to investigate the non-linear $\sigma$-model further we write
the matrix ${\hat Q}$ in the form
\begin{equation}\label{SM6}
{\hat Q}=\left(
\begin{array}{cc}
\sqrt{1-q_{12}q_{21}}&q_{12}\\
q_{21}&-\sqrt{1-q_{21}q_{12}}
\end{array}
\right),
\end{equation}
where
\begin{equation}\label{SM7}
q_{12}\equiv {q_{\alpha\alpha'}}_{\lambda}^{\kappa}\sigma_{\kappa}
\otimes\sigma_{\lambda},
\end{equation}
$\alpha>0$, $\alpha'\leq 0$ and $q_{21}=q_{12}^+$, where  + 
symbolizes  hermitian conjugation, and expand the 
action (\ref{SM3}) with respect to powers of $q$ up to fourth order
in $q$. Doing so, we use
the fact that the matrices ${q_{\alpha\alpha'}}_{\kappa}^{\lambda}$
are antisymmetric with respect to  $\alpha$ and $\alpha'$
for $\lambda=0,1,2$ and $\kappa=0$ and for $\lambda=3$ and $\kappa=1,2,3$
and symmetric with respect to $\alpha$ and $\alpha'$
for $\lambda=3$ and $\kappa=0$ and for $\lambda=0,1,2$ and $\kappa=1,2,3$.
The quadratic terms in the expansion yield the Gaussian propagator, the
physics of which agrees with the ladder-approximation.
The quartic terms in the expansion yield the weak-localization
corrections to the transport coefficients.
\subsection{Diffusion and relaxation in the Gaussian approximation}
In the Gaussian approximation the effective action takes the form
%
%
\begin{eqnarray}\label{GA1}
A_{\sigma}^{(2)}=-\pi\nu \hspace{-0.5em}&\hbar& 
\hspace{-0.5em}\sum_{{\alpha>0}\atop{\alpha'\leq 0}}
\sum_{\lambda,\kappa,\delta}
\int d{\bm x}\theta(\mu_x)\nonumber\\
&\times&{q_{\alpha\alpha'}}_{\lambda}^{\kappa}({\bm x})
\Gamma_{\lambda\delta}({\bm x}|E,s)\eta_{\delta}^{\kappa}
{q_{\alpha\alpha'}}_{\delta}^{\kappa}({\bm x}),
\end{eqnarray}
%
%
where $\eta_{\lambda}^{\kappa}=1$ for $\lambda=0,1,2$ and $\kappa=1,2,3$
and for $\lambda=3$ and  $\kappa=0$ and $\eta_{\lambda}^{\kappa}=-1$ else.
The operator $\Gamma_{\lambda\delta}({\bm x}|E,s)$, the non-vanishing 
components of which are
\begin{equation}\label{GA2}
\Gamma_{00}({\bm x}|E,s)=s-({\bm\nabla},D(\mu_x){\bm\nabla})
\end{equation}
and
\begin{eqnarray}\label{GA3}
\Gamma_{ik}({\bm x}|E,s)\hspace{-0.5em}&=&\hspace{-0.5em}
(s+\Omega_i(\mu_x)(1+\delta_{i3})-
({\bm\nabla},D_i(\mu_x){\bm\nabla}))\delta_{ik}\nonumber\\
&+&\hspace{-0.5em}\frac{N_z}{2}
(\delta_{k3}\{\omega_s(\mu_x),\nabla_i\}-\{\omega_s(\mu_x),\nabla_i\}
\delta_{i3}),\nonumber\\
\end{eqnarray}
is related to the 
generalized
diffusion propagator $P_{\lambda\kappa}({\bm x},{\bm x'}|E,s)$ by the 
relationship
\begin{equation}\label{GA4}
\sum_{\lambda}\Gamma_{\kappa\lambda}({\bm x}|E,s)P_{\lambda\delta}({\bm x},
{\bm x'}|E,s)=\delta_{\kappa\delta}\delta({\bm x}-{\bm x'}).
\end{equation}
Here the bracket $\{..,..\}$ symbolizes the anti-commutator,
$\hbar s=\omega/2$ and the indices $i$ and $k$ can take on the values $1,2,3$, which 
correspond to $x$, $y$, and $z$, respectively. The transport coefficients are 
given by
\begin{equation}\label{GA5}
\Omega_i(\mu_x)=\frac{4N_z^2m^2}{\hbar^2}D(\mu_x),
\end{equation}
\begin{equation}
\omega_s(\mu_x)=\frac{4m}{\hbar}D(\mu_x),
\end{equation}
and $D_i(\mu_x)=D(\mu_x)$. 

According to the Eqs.(\ref{GA2}), (\ref{GA3}) and (\ref{GA4})
the equation for   the particle transport decouples 
from the equations for the spin transport. The particle transport is 
described by the propagator $P_{00}$. Its equation of motion contains 
besides ordinary particle diffusion also the particle packet drift and 
the heating of the charge carriers. Its physics is detailed in 
Ref.[\onlinecite{Bleibaum1}]. The spatial components of the diffusion 
propagator,  $P_{ik}$ for $i,k=1,2,3$, describe spin diffusion and 
relaxation. In addition to the diffusion, drift and heating processes
present in the particle number diffusion equation, these equations also
contain the spin relaxation and an electric-field dependent rotation of
the spin-direction, which is induced by the Rashba-interaction. We would like
to note that the same equations can also be obtained from the ladder 
approximation.
\subsection{Quantum corrections}
In order to calculate the quantum corrections to the transport
coefficients we also
take into account the quartic terms in the expansion of $A_\sigma$
with respect to $q$ and investigate the corresponding functional 
integral in one-loop approximation. Doing so, we restrict the consideration 
to the linear response regime. In this case we can ignore the impact of 
the electric field on the transport coefficients and thus set the external 
electric field ${\bm F}$ equal to zero in all loops. For charge transport
the range of validity of this approximation has been discussed in 
Ref.[\onlinecite{Bleibaum1}]. There it has been shown that the impact of 
the field is negligible, as long as the spread of the particle packet
is small compared to the distance of its center from the turning point.
The time needed by a particle packet, which was a delta-pulse
centered at ${\bm x}={\bm 0}$ at time $t=0$ and in which all particles 
have the same energy $E$, to reach a width of the order of $E/F$ is for a 
typical experiment larger than $10^{-4}$ s, as shown in 
Ref.[\onlinecite{Bleibaum1}]. Given that the typical lifetime
of an initial magnetization is in the nanosecond range it seems to be 
reasonable to restrict the consideration to this limit.    

To one-loop order the particle number diffusion coefficient $D(E)$ 
in Eq.(\ref{GA2}) is replaced by  
\begin{eqnarray}\label{WL1}
D^{(1)}(E,s)=D(E)(1-\frac{1}{2\pi\nu\hbar}
\hspace{-0.5em}&(&\hspace{-0.5em}P_{33}(E,s)+2P_{11}(E,s)\nonumber\\
& &-P_{00}(E,s))),
\end{eqnarray}
where
\begin{equation}\label{WL2}
P_{00}(E,s)=\frac{1}{4\pi D(E)}\ln\frac{D(E){\tilde\Lambda}^2}{s},
\end{equation}
\begin{equation}\label{WL3}
P_{11}(E,s)=\frac{1}{4\pi D(E)}\ln\frac{D(E){\tilde\Lambda}^2}{s+\Omega(E)},
\end{equation}
and
\begin{equation}\label{WL4}
P_{33}(E,s)=\frac{1}{4\pi D(E)}\ln\frac{D(E){\tilde\Lambda}^2}
{\sqrt{s^2+2\Omega^2(E)+3s\Omega(E)}}.
\end{equation}
Here ${\tilde\Lambda}$ is the ultraviolet cutoff. These corrections have the
same structure as the conventional anti-weak localization corrections
to the conductivity \cite{Larkin}. In a current relaxation experiment
the frequency $s$ is effectively replaced by  
$1/\tau_{\phi}(T)$, where $\tau_{\phi}(T)$ is the inelastic
phase breaking time which increases with decreasing temperature. 
Accordingly,  the particle number diffusion 
coefficient has a minimum as a function of the temperature at 
$T=T_{\Omega}$, where $\Omega(E)\approx 1/\tau_{\phi}(T_{\Omega})$. 
Therefore, the diffusion coefficient increases with decreasing 
temperature if $T<T_{\Omega}$. 

The opposite turns out to be true  for the spin-transport 
coefficients. To one-loop order the spin diffusion coefficients 
$D_x(E)$ and $D_y(E)$ are replaced by
\begin{equation}\label{WL5}
D_{x/y}^{(1)}(E,s)=D(E)(1-\frac{1}{2\pi\nu\hbar}(P_{00}(E,s)+P_{33}(E,s))),
\end{equation}
$D_z(E)$ by
\begin{eqnarray}\label{WL6}
D_z^{(1)}(E,s)=D(E)(1-\frac{1}{2\pi\nu\hbar}
\hspace{-0.5em}&(&\hspace{-0.5em}
P_{00}(E,s)+2P_{11}(E,s)\nonumber\\
& &-P_{33}(E,s))),
\end{eqnarray}
the spin relaxation frequencies $\Omega_i(E)$ by
\begin{equation}\label{WL7}
\Omega_{x/y}(E,s)=\Omega(E)\frac{D^{(1)}_z(E,s)}{D(E)}
\end{equation}
and
\begin{equation}\label{WL8}
\Omega_z^{(1)}(E,s)=\Omega(E)\frac{D^{(1)}_x(E,s)}{D(E)},
\end{equation}    
and the rotation frequency $\omega_s(E)$ by
\begin{equation}\label{WL9}
\omega_s^{(1)}(E,s)=\omega_s(E)(1-\frac{1}{2\pi\nu\hbar}
(P_{00}(E,s)+P_{11}(E,s))).
\end{equation}
In these equations $s$ has also to be replaced by $1/\tau_{\phi}(T)$
if $s$ is smaller than $1/\tau_{\phi}(T)$. In contrast to
the particle-number diffusion coefficient, however, the spin-transport 
coefficients 
have no  minimum as a function of the temperature. In the range of
applicability of the Eqs.(\ref{WL5})-(\ref{WL9}) the magnitude of the
spin transport coefficients keeps  decreasing with decreasing temperature 
even if $T<T_{\Omega}$. In contrast to the corrections to the particle-
number diffusion coefficient,  which are anti-localizing at low 
temperatures, the corrections to the spin transport coefficients always
show a tendency to localization. 

At this point we would also like to mention that the quantum corrections
render the spin-transport coefficients weakly anisotropic. This anisotropy 
results from the fact that $P_{11}(E,s)\neq P_{33}(E,s)$. However, since
the difference between these quantities is of the order of $1/4\pi\nu\hbar
D(E)$ this difference is small compared to the magnitude of the quantum 
corrections themselves, which are larger than the former by a factor of 
the order of $\ln (D(E){\tilde\Lambda}^2/(s+\Omega(E)))$. Therefore,
we ignore the anisotropy in the following and replace $P_{33}(E,s)$ by
$P_{11}(E,s)$ everywhere. In this case $\Omega^{(1)}_{x}(E,s)
=\Omega^{(1)}_{z}(E,s)=\Omega^{(1)}(E,s)$ and $D^{(1)}_{x}(E,s)
=D^{(1)}_{z}(E,s)=D_s^{(1)}(E,s)$
\section{Application to relaxation phenomena}
The evolution of an initial spin packet is described by the Bloch equations.
Phenomenology tells us that the Bloch equations have
the structure
\begin{equation}\label{DMM1}
s{\bm S}(s)+{\bm \Omega}(s) \cdot {\bm S}(s)+{\bm R}(s)\times {\bm S}(s)
+{\mbox{div}}
{\bm J}(s)={\bm S}_0.
\end{equation} 
Here ${\bm S}(s)$ is the Laplace transform of the spin density with respect 
to time, $s$ is the corresponding Laplace frequency, ${\bm S}_0$ is the
initial condition, ${\bm\Omega}(s)$ is a
second rank tensor, ${\bm R}(s)$ is a vector, and 
${\bm J}(s)$ is the spin-current tensor. For the components we use the
notation  $({\bm \Omega}\cdot {\bm S})_i=\sum_k\Omega_{ik}S_k$ and 
$({\mbox{div}{\bm J}})_i=\sum_k\nabla_k J_{ki}$. The quantities ${\bm S}(s)$,
${\bm S}_0$ and ${\bm J}(s)$ depend also on ${\bm x}$ and on the total energy 
$E$, but  this dependence has been suppressed in 
Eq.(\ref{DMM1}) to simplify the notation. Due to the generalized translation 
invariance (\ref{B4}) the transport coefficients
${\bm{\Omega}}$ and ${\bm R}$ do not depend on the
quantities ${\bm x}$ and $E$ separately, but only on the kinetic energy 
$\mu_x$. In Eq.(\ref{DMM1}) this dependence has also been suppressed 
to simplify the notation.

From the physical point of view the tensor ${\bm{\Omega}}(s)$ describes the 
decay of the initial magnetization and ${\bm R}(s)$ describes the rotation of 
the initial magnetization. The disorder renders both quantities dispersive.
Therefore, both ${\bm \Omega}(s)$ and ${\bm R}(s)$ depend on $s$. This 
dependence should be observable in optical experiments where it should 
reflect itself in deviations from the purely exponential decay.

The equations (\ref{GA3}) and (\ref{GA4}) can be cast into the form
(\ref{DMM1}). Doing so, it has to be taken into account that the bare
transport coefficients are renormalized by the one-loop corrections
of the preceding section.From Eq.(\ref{GA3}) we find
\begin{equation}\label{DMM2}
\Omega_{ik}(\mu_x,s)=\Omega^{(1)}(\mu_x,s)\delta_{ik}(1+\delta_{i3}),
\end{equation}
and
\begin{equation}\label{DMM3}
{\bm R}(\mu_x,s)=\frac{1}{2}\frac{d\omega_s^{(1)}(\mu_x,s)}{d\mu_x}({\bm F}
\times{\bm N}).
\end{equation}
The spin current tensor takes the form
\begin{widetext}
\begin{equation}\label{DMM4}
{J}_{ki}({\bm x}|E,s)=-D_s^{(1)}(\mu_x,s)\nabla_kS_i({\bm x}|E,s)
+\omega_s^{(1)}(\mu_x,s)(N_iS_k({\bm x}|E,s)-\delta_{ik}({\bm N}\cdot
{\bm S}({\bm x}|E,s))).
\end{equation}
\end{widetext}
This tensor has been obtained by collecting all those terms in 
Eq.(\ref{GA3}), which can be cast into a divergence. Accordingly,
the tensor (\ref{DMM4}) might deviate from the true spin-current 
tensor by terms which are annihilated by the divergence. However,
since every observable quantity can be calculated directly from 
Eq.(\ref{DMM1}) such terms are not important for our further
considerations.

The Eqs.(\ref{WL1})-(\ref{WL9}) ignore  the impact of 
inelastic scattering events on the transport coefficients. To take 
them into account we introduce the phenomenological
phase relaxation time $\tau_{\phi}(T)$ and replace 
$s\to s+1/\tau_{\phi}(T)$. In this case the transport
coefficients are given by the Eqs.(\ref{DMM2}) 
and (\ref{DMM3}) with $s$ replaced by $1/\tau_{\phi}(T)$  for 
$s\ll 1/\tau_{\phi}(T)$. 
Accordingly, the relaxation becomes Markovian for $t>\tau_{\phi}(T)$.
Below we restrict the consideration to this situation.



\subsection{Decay of a homogeneous initial magnetization}
In order to investigate some of the consequences of Eq.(\ref{DMM1})
we first focus on the decay of a spatially homogeneous 
magnetization. To this end we assume that an 
initial magnetization ${\bm S}_0$ has been created at time $t=0$
and that only charge carriers in the vicinity of the Fermi-energy 
$\mu$ contribute to the magnetization. In this case the 
initial  distribution function is given by
\begin{equation}\label{DM3}
{\bm S}^0|_{F=0}({\bm x},E)={\bm S}_0\delta(\mu-E),
\end{equation}
in the absence of the field. In the presence of the field 
the energy of a charge carrier placed at ${\bm x}$ is changed by 
${\bm F}\cdot{\bm x}$. Therefore, the initial condition takes the 
form 
\begin{eqnarray}\label{DM4}
{\bm S}^0({\bm x},E)&=&{\bm S}_0\delta(\mu+{\bm F}\cdot{\bm x}-E)
\nonumber\\
&=&{\bm S}_0\delta(\mu-\mu_x)
\end{eqnarray}
in the presence of the field. 

During the evolution the spin distribution remains spatially homogeneous.
Accordingly,
${\bm S}({\bm x}|E,s)={\bm S}(\mu_x,s)$. The direction 
and the magnitude of the magnetization and the energy of the particles are 
changed. Since, however, the magnetization decays on a time scale  
small compared to the time needed by a particle to change its energy 
appreciably the energy distribution function remains close to the 
initial delta-function during the time in which the magnetization
is measurable. To take advantage of 
this fact we make the ansatz
\begin{equation}\label{DM5}
{\bm S}(\mu_x,s)={\bm S}(s)\delta(\mu-\mu_x)
\end{equation}
and derive a closed equation for ${\bm S}(s)$ by integrating Eq.
(\ref{DMM1}) with respect to $\mu_x$. Doing so, we obtain
the simple algebraic system of equations
\begin{equation}\label{DM6}
(s+\Omega^{(1)}(\mu))S_x(s)-\frac{1}{2}\frac{d\omega_s^{(1)}(\mu)}{d\mu}
[({\bm N}\times{\bm F})\times{\bm S}(s)]_{x/y}=S^0_{x/y},
\end{equation}
\begin{equation}\label{DM7}
(s+2\Omega^{(1)}(\mu))S_z(s)
-\frac{1}{2}\frac{d\omega_s^{(1)}(\mu)}{d\mu}
[({\bm N}\times{\bm F})\times{\bm S}(s)]_{z}=S^0_{z}.
\end{equation}
This system  
has the same structure as that for the decay of an initial magnetization 
in an ordered system (hopping transport of
small polarons on an ordered lattice) \cite{Damker}. 
Only the structure of the transport coefficients is different.

\begin{figure}[t]
\includegraphics[width=9.5cm,bb=0 250 586 742]{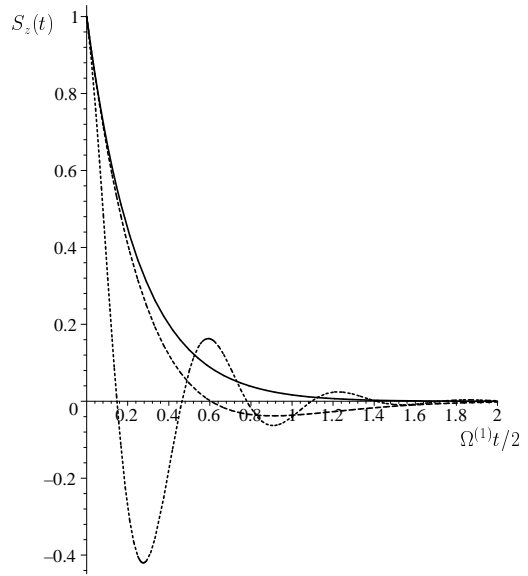}
\caption{Decay of $S_z$ for $2\Omega_F/\Omega^{(1)}=0.3$ (solid line), 
2 (dashed line) and 10 (dashed dotted line). The initial conditions
are $S_z(0)=1$ and $S_x(0)=0$.}
\end{figure}
\begin{figure}[t]
\includegraphics[width=9.5cm,bb=0 250 586 742]{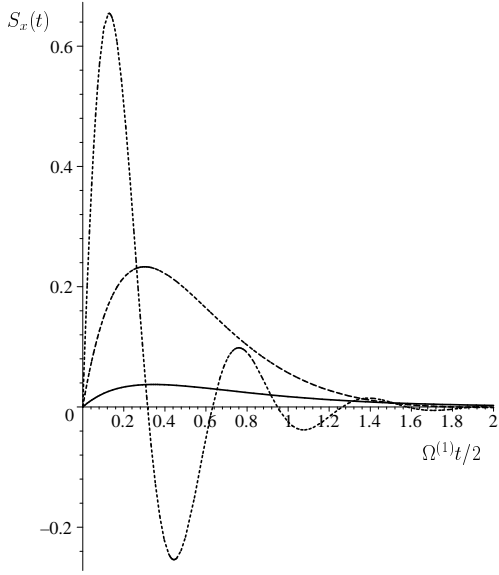}
\caption{$S_x(t)$ for $S_z(0)=1$ and $S_x(0)=0$. The parameters
are the same as in Fig.1.}
\end{figure}
To calculate the magnetization we solve this system and perform
an inverse Laplace transformation. Doing so, we obtain
\begin{equation}\label{DM8}
S_y(t)=S_y^0 e^{-\Omega^{(1)}t},
\end{equation}
%
%
\begin{eqnarray}\label{DM9}
S_x(t)=e^{-3\Omega^{(1)}t/2}
&[&S_x^0[\cosh(\omega_F t)+\frac{\Omega^{(1)}}{2}
\frac{\sinh(\omega_F t)}
{\omega_F}]\nonumber\\
& &+
S_z^0\Omega_F\frac{\sinh(\omega_F t)}
{\omega_F}]
\end{eqnarray}
and
\begin{eqnarray}\label{DM10}
S_z(t)=e^{-3\Omega^{(1)}t/2}
&[&S_z^0[\cosh(\omega_F t)-\frac{\Omega^{(1)}}{2}
\frac{\sinh(\omega_F t)}
{\omega_F}]\nonumber\\
& &-
S_x^0\Omega_F\frac{\sinh(\omega_F t)}
{\omega_F}].
\end{eqnarray}
%
%
Here $\Omega^{(1)}\equiv\Omega^{(1)}(\mu)$,
\begin{equation}\label{DM10a}
\omega_F=\sqrt{{\Omega^{(1)}}^2-
4\Omega_F^2}/2,
\end{equation}
and 
\begin{equation}\label{DM11}
\Omega_F=\frac{1}{2}\frac{d\omega_s(\mu)}{d\mu}FN_z.
\end{equation}
The Eqs.(\ref{DM8})-(\ref{DM10}) apply for $t>\tau_{\phi}(T)$. They do
not contain the magneto-electric effect of Ref.
[\onlinecite{Edelstein}], since this effect is proportional to the
ratio between the Rashba level splitting and the Fermi energy, which is 
considered as small in our treatment.

According to the Eqs.(\ref{DM8})-(\ref{DM11}) the character of the 
decay of the initial magnetization depends on the magnitude of the
electric field. If $F<F_c$, where $F_c$ is determined from the
requirement $\Omega^{(1)}=2|\Omega_{F_c}|$, the frequency $\omega_F$ is 
real and reflects itself in an exponential decay of an initial 
magnetization (see the Figs.1 and 2). Furthermore, the $x$-component of 
the magnetization is coupled with the $z$-component of the magnetization. 
If $S_z^0\neq 0$ a non vanishing magnetization in $x$-direction is created  
 even if $S_x^0=0$. If  $F>F_c$
the frequency $\omega_F$ becomes purely imaginary. In this case the
hyperbolic functions turn to trigonometric functions, which 
lead to a rotation of the decaying initial magnetization (see Fig.1 
and 2). Therefore, the  vector $({ S}_x(t), S_{z}(t))$, with components
\begin{eqnarray}\label{DM99a}
S_x(t)=e^{-3\Omega^{(1)}t/2}
&[&S_x^0[\cos(|\omega_F| t)+\frac{\Omega^{(1)}}{2}
\frac{\sin(|\omega_F |t)}
{|\omega_F|}]\nonumber\\
& &+
S_z^0\Omega_F\frac{\sin(|\omega_F| t)}
{|\omega_F|}]
\end{eqnarray}
and
\begin{eqnarray}\label{DM10aa}
S_z(t)=e^{-3\Omega^{(1)}t/2}
&[&S_z^0[\cos(|\omega_F| t)-\frac{\Omega^{(1)}}{2}
\frac{\sin(|\omega_F| t)}
{|\omega_F|}]\nonumber\\
& &-
S_x^0\Omega_F\frac{\sin(|\omega_F| t)}
{|\omega_F|}],
\end{eqnarray}
moves on
an  ellipse with decaying diameter. The ellipse satisfies the equation
\begin{equation}\label{DM13}
\Omega_F{{S_x}}^2(t)+\Omega{ S}_x(t){S}_z(t)+
\Omega_F{S}^2_z(t)=\mbox{const}\cdot\exp(-3\Omega^{(1)} t)
\end{equation}
The quadratic form can be diagonalized. Doing so,  it can be shown that
the elliptic axis are tilted by $\pi/4$. 
\subsection{Alternating Hall-current}
The rotation of the magnetization discussed in the preceding subsection
reflects also in the current. In the presence of the Rashba interaction 
the current operator has the form\cite{Molenkamp2,Schroll}
\begin{equation}\label{AH1}
{{\hat{\bm j}}}(t)=e\frac{{\hat{{\bm p}}}(t)}m-\frac{2e}{\hbar}
{{\hat{\bm S}}}(t)\times
{\bm{N}}.
\end{equation}
Here $e$ is the charge and ${\hat{\bm{S}}}$ is the spin operator. The 
Heisenberg equation for ${\hat{\bm{p}}}(t)$ is independent of 
${\hat{\bm{S}}}(t)$. Accordingly, the configuration average of the
expection value of the first term on the right hand side of 
Eq.(\ref{AH1}) yields the conventional Drude current and its weak
localization corrections. The configuration
average of the second term yields two additional 
contributions, a correction term to the longitudinal current
\begin{equation}\label{AH2}
 \delta j_x(t)=-\frac{2e}{\hbar}N_zS_y(t),
\end{equation}
and the Hall-like term
\begin{equation}\label{AH3}
j_y(t)=\frac{2e}{\hbar}N_zS_z(t)
\end{equation}
perpendicular to both ${\bm F}$ and ${\bm {N}}$. If we chose
the same initial condition for the magnetization as in the preceding 
section $S_z(t)$ is oscillating, if $2|\Omega_F|>\Omega^{(1)}$. According
to Eq.(\ref{AH3}) these oscillations manifest themselves in an 
alternating Hall-like current.The alternating Hall current is produced
by a constant electric field. However, since the magnetization decays when 
time elapses the current oscillations are not stable but decay too. This
raises the interesting question whether these oscillations can be 
stabilized by some means.

\section{Results}  
In the paper we have used field theoretical methods to derive
the generalized Bloch equations for the magnetization in a disordered 
two-dimensional electron gas. Our equations apply for time scales 
which are large compared to the elastic scattering time, large compared
to the phase-breaking time but small compared to energy relaxation time.
Accordingly, they take into account the particle- and spin diffusion, 
the drift induced by an external electric field in  the plane, the heating 
of the charge carriers, the decay of the magnetization due to 
scattering events, and an electric field dependent rotation of the
magnetization direction. 

The calculation of the transport coefficients shows 
that the impact of the quantum-corrections on the spin transport coefficients 
is different from that on the particle transport coefficients. The
corrections to the latter are antilocalizing. This fact manifests itself
in a minimum of the diffusion coefficient as a function of the temperature.
The structure of the corrections to the particle transport coefficients
agrees with the conventional antilocalization corrections to the conductivity
due to spin-orbit coupling\cite{Larkin}.
In contrast to the particle transport coefficients the
spin transport coefficients show a tendency to localization. Both the
spin diffusion coefficient and the spin decay rate decrease with 
decreasing temperature, even at very low temperatures. This sets the situation
for the spin-transport coefficients apart from that for the particle transport
coefficients, which increase with decreasing temperature at sufficiently
low temperature due to antilocalization. 

We have applied these equations to an investigation of the decay
of a spatially homogeneous initial magnetization. In the absence of 
the in-plane electric field such a magnetization decays exponentially
at large times\cite{Bleibaum2}. The decay is anisotropic. Its transverse 
component
decays twice as fast as its longitudinal components.The temperature 
dependence of the decay is given by the temperature dependence of the
weak localization corrections to the transport coefficients, which
lead to a logarithmic reduction of the decay rate with decreasing 
temperature. Accordingly, the decay becomes slower with decreasing
temperature.

Already a weak electric field renders the decay qualitatively different.
If the electric field is smaller than a critical field it tips the
magnetization only slightly over. If, however, the field exceeds the
critical field $F_c$ the magnetization starts rotating  in the plane 
transverse to ${\bm N}\times{\bm F}$. The rotation frequency depends on 
the degree of disorder, on the magnitude of the Rashba-parameter, on the 
magnitude of the electric field and on temperature. An increase of the
temperature results also in an increase of the rotation frequency
$\omega_F$ and in an increase of the critical field $F_c$. We would like 
to note that a
similar rotation has also been  obtained in Ref.[\onlinecite{Damker}]
for small polarons on an ordered lattice. Our calculation shows that
this rotation can also be observed in a disordered electron gas, in which
the transport mechanism is very different from that for a small
polaron on an ordered lattice. 
\acknowledgments This work was done during a stay at the
the University of Oregon. The author is grateful for the  
hospitality of the material science 
institute, 
in particular to D. Belitz, without whose support and encouragement
this work would not have been 
accomplished. 
I would also like to thank V. V. Bryksin for helpful 
discussions. 
This work was supported by the DFG under grant. No.
Bl456/3-1. 


\begin{thebibliography}{}
\bibitem{Bleibaum2} The magneto-electric effect of 
Ref.[\onlinecite{Edelstein}] is absent in our approach since
we have ignored the Rashba level splitting compared
to the Fermi energy.
\end{thebibliography}
\end{document}